\documentclass[prl,aps,twocolumn,superscriptaddress]{revtex4}

\usepackage[dvips]{graphicx}
\usepackage{amssymb,amsfonts,amsmath}
\usepackage{color}
\usepackage{ulem}
\usepackage[hidelinks]{hyperref}
\usepackage{bbm}
\usepackage{bm}

\newcommand{\rv}{{\mathbf r}}

\newcommand{\Tr}{{\rm Tr}\,}

\newcommand{\pv}{{\bf p}}

\newcommand{\Fv}{{\bf F}}

\newcommand{\ms}{}

\newcommand{\mz}{\color{black}}

\newcommand{\eps}{{\boldsymbol \epsilon}}
\newcommand{\unity}{{\mathbbm 1}}
\newcommand{\cov}{{\rm cov}}

\newcommand{\ext}{{\rm ext}}

\newcommand{\avg}[1]{\Big\langle #1 \Big\rangle}
\newcommand{\eqr}[1]{Eq.~\eqref{#1}}
\newcommand{\ff}{{f\!f}}
\newcommand{\gradf}{{\nabla\!f}}

\newcommand{\mydelete}[1]{{}}

\begin{document}

\title{\mz Noether's Perspective on the Equilibrium Structure of Liquids}
\title{\mz Noether's Theorem Strongly Constrains Correlations in Equilibrium Liquids}
\title{\mz Noether-Constrained Correlations in Equilibrium Liquids}
\title{\mz Noether-Constrained Correlations in Equilibrium Liquids}

\author{Florian Samm\"uller}
\affiliation{Theoretische Physik II, Physikalisches Institut, 
  Universit{\"a}t Bayreuth, D-95447 Bayreuth, Germany}
\author{Sophie Hermann}
\affiliation{Theoretische Physik II, Physikalisches Institut, 
  Universit{\"a}t Bayreuth, D-95447 Bayreuth, Germany}
\author{Daniel de las Heras}
\affiliation{Theoretische Physik II, Physikalisches Institut, 
  Universit{\"a}t Bayreuth, D-95447 Bayreuth, Germany}
\author{Matthias Schmidt}
\affiliation{Theoretische Physik II, Physikalisches Institut, 
  Universit{\"a}t Bayreuth, D-95447 Bayreuth, Germany}
\email{Matthias.Schmidt@uni-bayreuth.de}

\date{26 January 2023; 
\mz revised version: 7 June 2023}

\begin{abstract}
Liquid structure carries deep imprints of an inherent thermal
invariance against a spatial transformation of the underlying
classical many-body Hamiltonian. At first order in the transformation
field Noether's theorem yields the local force balance. Three distinct
two-body correlation functions emerge at second order, namely the
standard two-body density, the localized force-force correlation
function, and the localized force gradient. An exact Noether sum rule
interrelates these correlators. Simulations of Lennard-Jones, {\ms
  Yukawa, soft-sphere dipolar, Stockmayer, Gay-Berne} and
Weeks-Chandler-Andersen liquids, of monatomic water and of a colloidal
gel former demonstrate the fundamental role in the characterization of
spatial structure.
\end{abstract}

\maketitle

It is a both surprising and intriguing phenomenon that the liquid
phase occurs {\ms in the phase diagram at and off coexistence} with
the gas or the solid phase. Famously it has been
argued~\cite{weisskopf1977,evans2019physicsToday} that it needs
relying on observations rather than mere theory alone to predict the
existence of {\ms liquids}, as neither the noninteracting ideal gas
nor the Einstein crystal form appropriate idealized references.  The
liquid state \cite{hansen2013, barker1976, evans1979, evans2016}
comprises high spatial symmetry against global translations and
rotations, together with the correlated and strongly interacting
behaviour of the dense constituents, whether they are atoms,
molecules, or colloids.

Among the defining features of {\ms liquids} are the ability to
spontaneously form an interface when at liquid-gas coexistence, the
viscous response against shearing motion, and the rich pair
correlation structure.  While the one-body density distribution is
homogeneous in bulk (in stark contrast to the microscopic density of a
crystal), the joint probability of finding two particles at a given
separation distance~$r$ is highly nontrivial in a liquid. The pair
correlation function $g(r)$ \cite{hansen2013}, as accessible e.g.\ via
microscopy \cite{royall2007gofr, thorneywork2014,
  statt2016,ramirez2006} {\ms and scattering \cite{hansen2013,
    yarnell1973, salmon2006,
carvalho2022, dyre2016topicalReview} techniques}, quantifies this
spatial structure on the particle level. At large distances $r$, the
asymptotic decay of $g(r)$ falls into different classes
\cite{evans1993decay,evans1994decay,dijkstra2000decay,
  grodon2004decay} with much current interest in
electrolytes~\cite{cats2021decay}. {\ms The spatial Fourier transform
  of $g(r)$ is the static structure factor \cite{hansen2013,
    yarnell1973, salmon2006, 
 carvalho2022, dyre2016topicalReview}.}

It is a common strategy to exploit the symmetries of a given physical
system via Noether's theorem of invariant variations
\cite{noether1918,byers1998}.  From symmetries in the dynamical
description of the system one systematically obtains conservation
laws. Typically the starting point is the action functional, as
generalized to a variety of statistical mechanical
settings~\cite{lezcano2018stochastic, baez2013markov,
  marvian2014quantum, sasa2016, sasa2019, sasa2020, 
  revzen1970, budkov2022, brandyshev2023}. In contrast, we have
recently applied Noether's concept directly to statistical mechanical
functionals, such as the free energy \cite{hermann2021noether,
  hermann2022topicalReview, hermann2022variance,
  hermann2022quantum}. This allows to exploit a specific thermal
invariance property of Hamiltonian many-body systems against shifting
as performed globally \cite{hermann2021noether,
  hermann2022topicalReview, hermann2022variance} or locally resolved
in position~\cite{hermann2022quantum, tschopp2022forceDFT}.

In this Letter we demonstrate that at the local second-order level the
thermal Noether invariance leads to exact identities {\ms (``sum
  rules'' \cite{hansen2013, henderson1984, henderson1985,
    henderson1992, evans1992, evans1990, upton1998, hirschfelder1960,
    triezenberg1972})} that form a comprehensive statistical two-body
correlation framework. We use simulations to demonstrate the relevance
for the investigation of the structure of simple, beyond-simple and
gelled liquids.

We consider systems of $N$ {\ms classical} particles {\ms in three
  dimensions} with positions $\rv_1,\ldots,\rv_N\equiv \rv^N$ and
momenta $\pv_1,\ldots,\pv_N\equiv \pv^N$. The Hamiltonian consists of
kinetic, interparticle, and external energy contributions,
\begin{align}
  H &= \sum_i \frac{\pv_i^2}{2m} + u(\rv^N) 
  + \sum_i V_{\rm ext}(\rv_i),
  \label{EQHamiltonian2}
\end{align}
where the indices $i=1,\ldots,N$ run over all particles, $m$~indicates
the particle mass, $u(\rv^N)$ is the interparticle interaction
potential, and $V_{\rm ext}(\rv)$ is a one-body external potential as
a function of position $\rv$.

We consider a canonical transformation \cite{goldstein2002}, where
coordinates and momenta change according to the following map
{\ms \cite{tschopp2022forceDFT}:}
\begin{align}
  \rv_i &\to \rv_i + \eps(\rv_i),
  \label{EQcanonicalTransformationCoordinates}\\
  \pv_i &\to [\unity + \nabla_i\eps(\rv_i)]^{-1}\cdot\pv_i.
  \label{EQcanonicalTransformationMomenta}
\end{align}
Here $\eps(\rv)$ is a spatial ``shifting'' field that parameterizes
the transform, $\unity$ indicates the $3\times 3$-unit matrix, the
superscript $-1$ of a matrix is its inverse, {\ms and $\nabla_i$
  indicates the derivative with respect to $\rv_i$, such that
  $\nabla_i\eps(\rv_i)$ is a $3\times 3$-matrix.} The transformation
\eqref{EQcanonicalTransformationCoordinates} and
\eqref{EQcanonicalTransformationMomenta} preserves both the phase
space volume element and the Hamiltonian~\cite{tschopp2022forceDFT,
  goldstein2002}; {\ms its self-adjoint version is applicable to
  quantum systems \cite{hermann2022quantum}. The form of the vector
  field $\eps(\rv)$ must be such that the transformation between
  original and new coordinates is bijective \cite{hermann2022quantum}.}

We consider the shifting field and its gradient to be small and hence
Taylor expand. The coordinate transformation
\eqref{EQcanonicalTransformationCoordinates} is already linear in the
displacement field and is hence unaffected.  The momentum
transformation~\eqref{EQcanonicalTransformationMomenta}, when expanded
as a geometric (Neumann) series to second order, is:
\begin{align}
  [\unity + \nabla_i\eps(\rv_i)]^{-1}
  =\unity - \nabla_i\eps(\rv_i)+
  [\nabla_i\eps(\rv_i)]^2-\ldots,
\end{align}
{\ms where the exponents on the right hand side imply matrix products
  such that $[\nabla_i\eps(\rv_i)]^2=[\nabla_i\eps(\rv_i)]\cdot
  [\nabla_i\eps(\rv_i)]$, {\it etc}.}
When expressed in the new variables, the Hamiltonian acquires a
functional dependence on the shifting field, i.e.\ $H\to H[\eps]$. It
is then straightforward to show
\cite{hermann2022quantum,tschopp2022forceDFT} that the locally
resolved one-body force operator $\hat\Fv(\rv)$ follows from
functional differentiation according to:
\begin{align}
  -\frac{\delta H[\eps]}{\delta \eps(\rv)}  \Big|_{\eps=0}
  &= 
  \hat\Fv(\rv),
  \label{EQFoperatorFromDifferentiation}
\end{align}
where $\delta/\delta\eps(\rv)$ indicates the functional derivative
with respect to the shifting field $\eps(\rv)$. As indicated,
$\eps(\rv)$ is set to zero after the derivative has been taken.
Similar to the structure of the Hamiltonian \eqref{EQHamiltonian2},
the one-body force operator $\hat\Fv(\rv)$ contains kinetic,
interparticle, and external contributions:
\begin{align}
  \hat\Fv(\rv) &=
  -\nabla\cdot\sum_i\frac{\pv_i\pv_i}{m}\delta(\rv-\rv_i)
  + \hat\Fv_{\rm int}(\rv)
  -\hat\rho(\rv)\nabla V_{\rm ext}(\rv).
  \label{EQForceDensityOperator}
\end{align}
Here $\delta(\cdot)$ indicates the (three-dimensional) Dirac
distribution, $\hat\Fv_{\rm
  int}(\rv)=-\sum_i\delta(\rv-\rv_i)\nabla_iu(\rv^N)$ is the
interparticle one-body force operator \cite{schmidt2022rmp}, and
$\hat\rho(\rv)=\sum_i\delta(\rv-\rv_i)$ is the standard one-body
density operator \cite{evans1979,hansen2013}. {\ms All considerations
  so far are general and hold per microstate.}

We complement this deterministic description by the statistical
mechanics of the grand ensemble at chemical potential $\mu$ and
temperature $T$.  The grand potential is $\Omega = -k_BT\ln\Xi$, with
the the grand partition sum $\Xi=\Tr {\rm e}^{-\beta(H-\mu N)}$.  Here
$k_B$ indicates the Boltzmann constant, $\beta=1/(k_BT)$ denotes
inverse temperature, and the classical ``trace'' operation in the
grand ensemble is given by $\Tr \cdot=\sum_{N=0}^\infty (N!
h^{3N})^{-1}\int d\rv_1\ldots d \rv_N \int d\pv_1\ldots d\pv_N \cdot$,
where $h$ denotes the Planck constant. The corresponding grand
probability distribution is $\Psi={\rm e}^{-\beta(H-\mu N)}/\Xi$ and
thermal averages are defined via $\langle\cdot\rangle=\Tr \Psi \cdot$,
as is standard. A primary example of a thermal average is the density
profile being the average of the one-body density operator,
i.e.\ $\rho(\rv)=\langle\hat\rho(\rv)\rangle$.

Via the transformed Hamiltonian $H[\eps]$, the grand partition sum
acquires functional dependence on the shifting field
\cite{hermann2022quantum,tschopp2022forceDFT}, i.e.\ $\Xi[\eps]$, and
so does the grand potential, i.e.\ $\Omega[\eps]$.{ Noether invariance
  \cite{hermann2021noether,hermann2022topicalReview} however implies
  that the grand potential does not change under the transformation,
  and hence
\begin{align}
  \Omega[\eps] = \Omega,
  \label{EQOmegaInvariant}
\end{align}
irrespectively of the form of $\eps(\rv)$.}  The first functional
derivative of \eqr{EQOmegaInvariant} with respect to the shifting
field~$\eps(\rv)$ then yields \cite{hermann2022quantum,
  tschopp2022forceDFT} the locally resolved equilibribum force density
balance relation $\Fv(\rv)=\langle\hat\Fv(\rv)\rangle=0$
\cite{hansen2013, schmidt2022rmp}.

Here we work at the second-order level and hence consider the second
derivative of \eqr{EQOmegaInvariant}, which yields
\begin{align}
  \frac{\delta^2\Omega[\eps]}{\delta\eps(\rv)\delta\eps(\rv')}
  \Big|_{\eps=0} &= 0.
  \label{EQomegaSecondDerivative1}
\end{align}
Evaluating the functional derivative on the left hand side~gives
\begin{align}
  \frac{\delta^2\Omega[\eps]}{\delta\eps(\rv)\delta\eps(\rv')}  &=
  -\beta\,\cov\Big(
  \frac{\delta H[\eps]}{\delta\eps(\rv)},
  \frac{\delta H[\eps]}{\delta\eps(\rv')}
  \Big)+
  \Big\langle
  \frac{\delta^2 H[\eps]}{\delta\eps(\rv)\delta\eps(\rv')}
  \Big\rangle,
  \label{EQomegaSecondDerivative2}
\end{align}
where the covariance of two observables (phase space functions) $\hat
A$ and $\hat B$ is defined in the standard way as $\cov(\hat A,\hat
B)=\langle \hat A\hat B\rangle- \langle\hat A\rangle \langle \hat
B\rangle$.  Rewriting the derivative $\delta H[\eps]/\delta\eps(\rv)$
as the negative force density operator via
\eqr{EQFoperatorFromDifferentiation}, inserting
\eqr{EQomegaSecondDerivative2} into \eqr{EQomegaSecondDerivative1},
and re-arranging gives the following locally resolved two-body Noether
sum rule:
\begin{align}
  \beta\langle \hat\Fv(\rv)\hat\Fv(\rv')  \rangle
  &=
  \Big\langle
  \frac{\delta^2 H[\eps]}{\delta\eps(\rv)\delta\eps(\rv')}
  \Big\rangle \Big|_{\eps=0}.
  \label{EQsumRule2generic}
\end{align}
We have replaced $\cov(\hat\Fv(\rv),\hat\Fv(\rv'))= \langle
\hat\Fv(\rv)\hat\Fv(\rv') \rangle$, because
$\langle\hat\Fv(\rv)\rangle=0$ in equilibrium
\cite{hansen2013,schmidt2022rmp}. The sum rule
\eqref{EQsumRule2generic} relates the force-force correlations at two
different positions (left hand side) with the mean curvature of the
Hamiltonian with respect to variation in the shifting field (right
hand side).  That such physically meaningful averages are related to
each other, at all positions $\rv$ and~$\rv'$, is highly nontrivial.

\begin{figure}
  \includegraphics[width=0.9\linewidth]{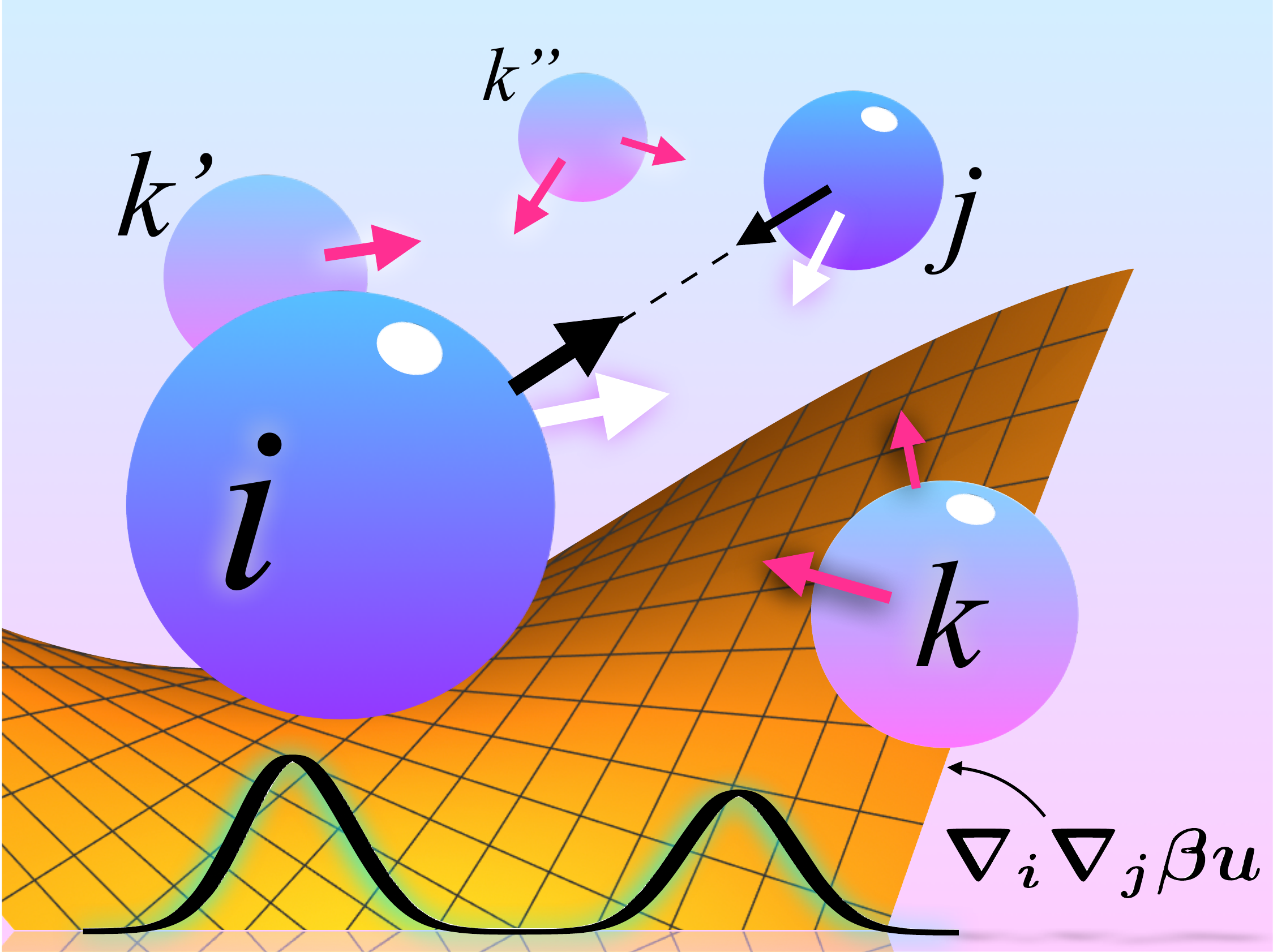}
  \caption{Illustration of the three different correlation functions
    that are constrained by thermal Noether invariance. The particles
    (spheres) exert forces (arrows) onto each other. Particles $i$ and
    $j$ interact directly with each other (black arrows).  The total
    force (white arrow) on each particle is also determined by the
    forces that all other particles~$k, k',k''$ exert (pink
    arrows). The force-force correlations are balanced by the
    potential energy curvature $\nabla_i\nabla_j \beta u(\rv^N)$
    (orange surface) and by the two-body density Hessian
    $\nabla\nabla'\rho_2(\rv,\rv')$ (black curve).
  \label{FIGillustration}}
\end{figure}

We can bring the fundamental Noether two-body sum rule
\eqref{EQsumRule2generic} into a more convenient form by multiplying
by~$\beta$, splitting off the trivial kinetic contributions, and
introducing the potential energy force operator $\hat\Fv_U(\rv)$,
which combines interparticle and external forces according to
$\hat\Fv_U(\rv) = \hat\Fv_{\rm int}(\rv)-\hat\rho(\rv)\nabla
V_\ext(\rv)$. Furthermore we focus on the distinct contributions
(subscript~``dist'') such that only pairs of particles with unequal
indices are involved and double sums reduce to
$\sum_{ij(\neq)}\equiv\sum_{i=1}^N\sum_{j=1,j\neq i}^N$. This allows
to identify from \eqr{EQsumRule2generic} the following exact distinct
two-body Noether identity:
\begin{align}
  &\avg{\beta\hat\Fv_U(\rv)\beta\hat\Fv_U(\rv')}_{\!\rm dist} =
  \nabla\nabla'\rho_2(\rv,\rv')
    \label{EQsumRuleFeFeDistinct}\\&\quad\qquad\qquad\quad
  + \avg{{\sum_{ij(\neq)}^{\phantom{space}}} 
    \delta(\rv-\rv_i)\delta(\rv'-\rv_j)
  \nabla_i\nabla_j \beta u(\rv^N)}.
  \notag
\end{align}
Here the two-body density is defined as is standard:
$\rho_2(\rv,\rv')=\langle\hat\rho(\rv)\hat\rho(\rv')\rangle_{\rm
  dist}=\langle\sum_{ij(\neq)}\delta(\rv-\rv_i)\delta(\rv'-\rv_j)\rangle$.
The relationship between the different correlators, as graphically
illustrated in Fig.~\ref{FIGillustration}, {\ms holds in general
  inhomogeneous situations with no need for specific simplifying
  symmetries.}

\begin{figure*}
\includegraphics[width=0.95\linewidth]{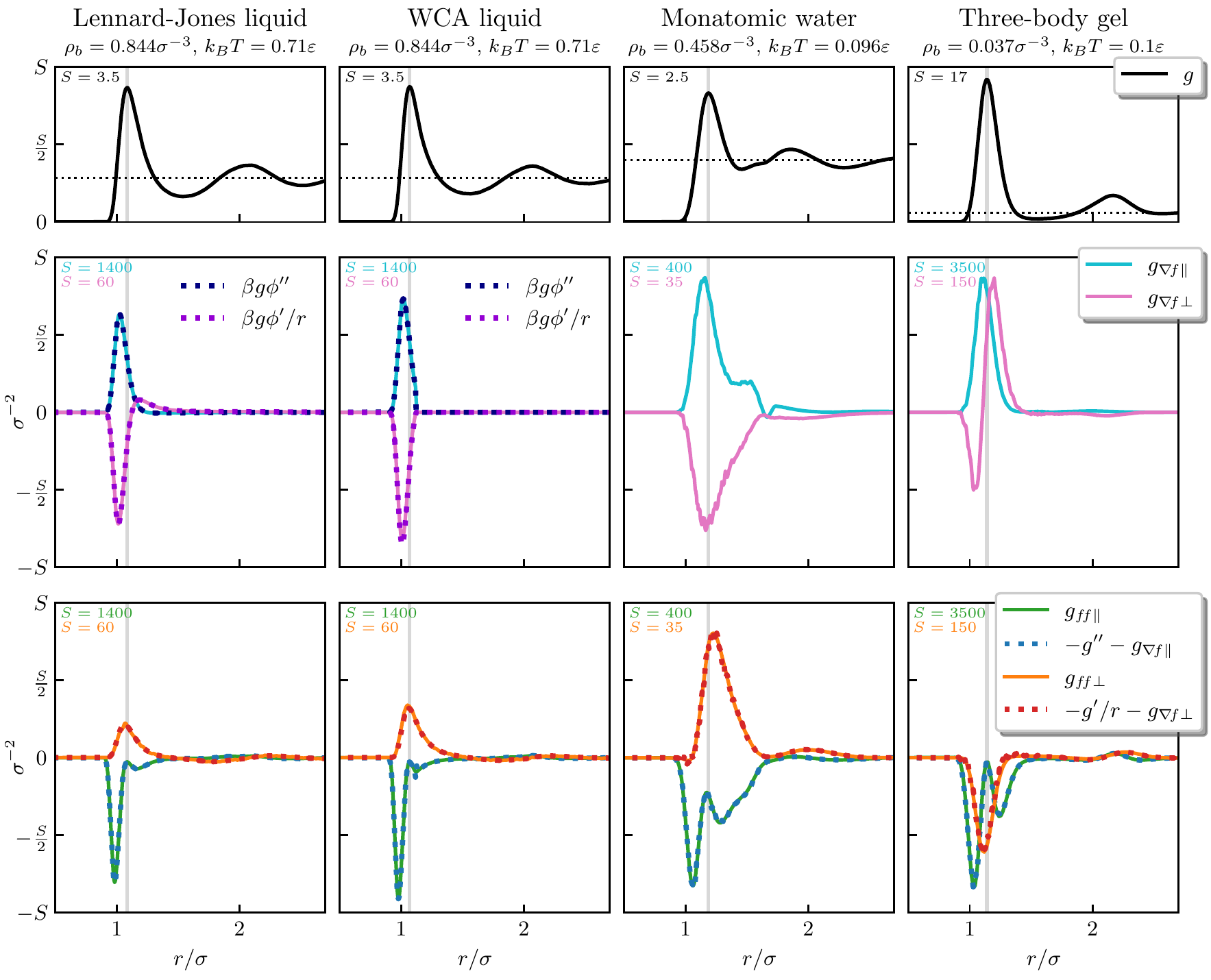}
\caption{Simulation results for the two-body correlation functions of
  the Lennard-Jones liquid (first column), the WCA liquid (second
  column), monatomic water (third column), and the three-body gel
  (fourth column). Results are shown as a function of the scaled
  interparticle distance~$r/\sigma$, $S$ is a vertical scale factor
  {\ms given in the upper left corner of each panel, and $\epsilon$
    denotes the energy scale of the respective model fluid.}  Shown is
  the pair correlation function $g(r)$ (top row), the potential
  curvature correlator ${\sf g}_{\gradf}(r)$ (middle row) and the
  force-force correlator ${\sf g}_{f\!f}(r)$ (bottom row); the latter
  two correlators have a transversal ($\perp$) and a parallel
  ($\parallel$) tensor component. The results for ${\sf g}_\gradf(r)$
  for the LJ and WCA liquids are numerically identical to those from
  the analytical expressions \eqref{EQsimpleFluidBoth} (dashed
  lines). The directly sampled results for ${\sf g}_\ff(r)$ are
  numerically identical to those obtained from the Noether sum rules
  \eqref{EQgIdentityManyBodyParallel} and
  \eqref{EQgIdentityManyBodyPerpendicular} (dashed lines) for all four
  systems. Vertical gray lines indicate the position of the first
  maximum of $g(r)$ as a guide to the eye.
  \label{FIGcorrelators}
}
\end{figure*}

We demonstrate that this framework has profound implications already
for a bulk liquid, where $\rho(\rv)=\rho_b=\rm const$ and
$V_\ext(\rv)=0$, such that $\hat\Fv_U(\rv)=\hat\Fv_{\rm int}(\rv)$. In
view of the form of the distinct sum rule
\eqref{EQsumRuleFeFeDistinct}, we use the pair correlation function
$g(|\rv-\rv'|)=\rho_2(\rv,\rv') /\rho_b^2$, and introduce both the
force-force pair correlation function ${\sf g}_\ff(|\rv-\rv'|)=\beta^2
\langle\hat\Fv_{\rm int}(\rv)\hat\Fv_{\rm int}(\rv')\rangle_{\rm
  dist}/\rho_b^2$, and the force gradient correlator ${\sf
  g}_\gradf(|\rv-\rv'|)=-\langle
\sum_{ij(\neq)}\delta(\rv-\rv_i)\delta(\rv'-\rv_j)\nabla_i\nabla_j
\beta u(\rv^N) \rangle/\rho_b^2$, which is also the negative mean
potential curvature.  The identity \eqref{EQsumRuleFeFeDistinct} can
then be written succinctly as:
\begin{align}
  \nabla\nabla g(r) & + {\sf g}_\gradf(r) + {\sf g}_\ff(r) = 0,
  \label{EQgIdentity}
\end{align}
where $r=|\rv-\rv'|$ denotes the separation distance between the two
positions.  Both ${\sf g}_\ff(r)$ and ${\sf g}_\gradf(r)$ have tensor
rank two, i.e.\ they are $3\times 3$-matrices. 
{\ms Given the central role that $g(r)$ plays in the theory of liquids
  \cite{hansen2013}, \eqr{EQgIdentity} is highly remarkable as it
  allows to express $g(r)$ via spatial integration of two seemingly
  entirely different (force-gradient and force-force) correlators.}
Due to the rotational symmetry of the bulk liquid, the only nontrivial
tensor components are parallel ($\parallel$) and transversal ($\perp$)
to $\rv-\rv'$, such that \eqr{EQgIdentity} reduces to
\begin{align}
  g''(r) + g_{\gradf\parallel}(r) + g_{\ff\parallel}(r)   &= 0,
  \label{EQgIdentityManyBodyParallel}\\
   g'(r)/r + g_{\gradf\perp}(r) + g_{\ff\perp}(r)  &= 0,
  \label{EQgIdentityManyBodyPerpendicular}
\end{align}
with the prime(s) denoting the derivative(s) with respect to~$r$.  In
the chosen coordinate system the matrices are diagonal, ${\rm
  diag}(\parallel,\perp,\perp)$, with the first axis being parallel to
$\rv-\rv'$.  
{\ms For molecular liquids of particles with orientational degrees of
  freedom \cite{hansen2013, zhao2011, jeanmairet2013jpcl,
luukkonen2020} our theory, including
  Eqs.~\eqref{EQgIdentityManyBodyParallel} and
  \eqref{EQgIdentityManyBodyPerpendicular}, remains valid upon
  equilibrium orientational averaging.}

For simple fluids, where the particles interact mutually only via a
pair potential $\phi(r)$, the force gradient correlator reduces to
${\sf g}_\gradf(r)=\beta g(r)\nabla\nabla \phi(r)$ such that
\begin{align}
 g_{\gradf\parallel}(r) &= \beta g(r)\phi''(r),
\quad
 g_{\gradf\perp}(r) = \beta g(r)\phi'(r)/r.
  \label{EQsimpleFluidBoth}
\end{align}
This simplification is due to the reduction of the mixed derivative
$\nabla_i\nabla_j u(\rv^N)=\nabla_i\nabla_j
\sum_{kl(\neq)}\phi(|\rv_k-\rv_l|)/2= \nabla_i \nabla_j
\phi(|\rv_i-\rv_j|)$, for $i\neq j$. This allows to rewrite the
curvature correlator in Eqs.~\eqref{EQgIdentityManyBodyParallel} and
\eqref{EQgIdentityManyBodyPerpendicular}, which attain the form
$g''(r) + \beta\phi''(r) g(r) + g_{\ff\parallel}(r)=0$ and $g'(r)/r +
\beta\phi'(r) g(r)/r + g_{\ff\perp}(r)=0$.
{\ms In the gas phase} the validity can be analytically verified {\ms
  on the second virial level}, where $g(r)=\exp(-\beta\phi(r))$ and
the force-force correlations are due to the antiparallel direct forces
between a particle pair:
$g_{\ff\parallel}(r)=-g(r)[\beta\phi'(r)]^2$. Furthermore
$g_{\ff\perp}(r)=0$ due to the absence of a third particle at
$\rho_b\to 0$ that could mediate a transversal force.

We substantiate this Noether correlation framework with computer
simulations using adaptive Brownian dynamics \cite{sammueller2021},
which is an algorithm that is both fast and allows for tight control
of force evaluation errors. We {\ms first} investigate the
Lennard-Jones (LJ) liquid, {the purely repulsive
  Weeks-Chandler-Andersen (WCA) liquid}, monatomic water
\cite{molinero2009,coe2022water}, and a three-body colloidal gel
former \cite{saw2009,saw2011}. The results are summarized in
Fig.~\ref{FIGcorrelators}; {\ms the top line gives the respective
  values of $T$ and $\rho_b=N/V$ with box volume $V=(10\sigma)^3$; the
  LJ potential is truncated at $r/\sigma=2.5$ with $\sigma$ denoting
  the respective particle size}.  We first discuss the two simple
liquids. Both the LJ and the WCA liquid feature pair correlation
functions $g(r)$ that display the familiar strongly structured, damped
oscillatory form \cite{evans1993decay,evans1994decay,hansen2013}, with
a prominent first peak indicating a nearest neighbor correlation shell
and subsequent, increasingly washed out oscillations at larger
distances. In stark contrast, both the force-gradient (potential
curvature) correlator ${\sf g}_\gradf(r)$ and the force-force
correlator ${\sf g}_\ff(r)$ have very different forms than $g(r)$
itself.  The curvature correlator has very strongly localized positive
{\ms($\parallel$)} and negative {\ms($\perp$)} peaks near
$r=\sigma$. This feature is due to the strong first peak of $g(r)$
combined with the properties of $\phi'(r)$ and $\phi''(r)$, as is
evident via \eqr{EQsimpleFluidBoth}, which we find to be satisfied to
high numerical accuracy.  Our results confirm the expectation
\cite{evans1994decay} that $g(r)$ is hardly affected by interparticle
attraction. In contrast the force gradient $g_{\gradf\perp}(r)$ has a
clear and significant peak in the attractive region of the LJ
potential with no such feature occurring in the purely repulsive WCA
liquid.

The force-force correlator ${\sf g}_\ff(r)$ has a similar first peak
structure as the curvature correlator, but oscillations extend much
further out to larger distances $r$. Hence ${\sf g}_\ff(r)$ captures
also the indirect interactions that are mediated by surrounding
particles; we recall Fig.~\ref{FIGillustration}. The strong negative
double peak of the parallel component indicates anti-correlated force
orientations, which reflect the direct interactions between pairs of
particles.  Both tensor components of ${\sf g}_\ff(r)$ satisfy the
Noether sum rules \eqref{EQgIdentityManyBodyParallel} and
\eqref{EQgIdentityManyBodyPerpendicular} to excellent numerical
accuracy.

To go beyond simple liquids, we first turn to the monatomic water
model by Molinero and Moore \cite{molinero2009}, which includes
three-body interparticle interactions in $u(\rv^N)$ that generate the
tetrahedral coordination of liquid water. The monatomic water model
gives a surprisingly accurate description of the properties of real
water, see Ref.~\cite{coe2022water} for very recent work, while the
particles remain spherical and there is no necessity to explicitly
invoke molecular orientational degrees of freedom. Hence our framework
\eqref{EQgIdentityManyBodyParallel} and
\eqref{EQgIdentityManyBodyPerpendicular} applies.  The third column of
Fig.~\ref{FIGcorrelators} demonstrates at ambient conditions that
while the shape of $g(r)$ is similar to that in the LJ liquid, both
the potential energy curvature and the force-force correlator differ
markedly from those of the LJ model. Notably the shape of the double
negative peak {\ms of $g_{\ff\parallel}(r)$} differs and the sign of
$g_{\ff\perp}(r)$ does not turn negative for distances towards the
second shell, as is the case in the LJ liquid.  Consistently, the
magnitude of the $\parallel$ component is much larger than the $\perp$
component, as direct interparticle interactions are prominent in the
former, whereas mediation by third particles is required for the
latter.

The three-body gel former by Kob and coworkers~\cite{saw2009,saw2011}
alters the prefered angle of the three-body interaction term from
tetrahedral to stretched (we use 180 degrees
\cite{sammueller2022gel}). This change induces an affinity for the
formation of chains while retaining an ability for their branching and
thus the model forms networks in equilibrium. The results shown in the
fourth column of Fig.~\ref{FIGcorrelators} indicate markedly different
behaviour as compared to the above liquids. While $g(r)$ has the
generic long-range decay that one expects of network-forming systems,
both the curvature and the force-force correlator are much more
specific indicators. In particular we attribute the striking shape of
the transversal ($\perp$) tensor component to the network
connectivity.  Again the sum rules are satisfied to very good
numerical accuracy which we take i) as a demonstration that the gel
state is indeed equilibrated, which distinguishes this model
\cite{saw2009,saw2011} from genuine nonequilibrium gel formers, and
ii) as a confirmation of the fitness of the Noether correlators to
systematically quantify complex spatial structure formation.
{\ms This holds beyond the presented model fluids; see the
  Supplementary Information \cite{sammueller2023supplementary} for
  results for screened long-ranged interparticle forces of Yukawa
  type, as well as for dipolar \cite{klapp2005topRev,
    teixeira2000topRev, stevens1995, tavares1999,allen2019topRev},
  Stockmayer \cite{allen2019topRev}, and (isotropic and nematic)
  Gay-Berne fluids \cite{allen2019topRev,gay1981,brown1998}.
{\mz For the LJ model, we also contrast the behaviour in the liquid
  against both the gas and the crystal, where the identities
  \eqref{EQgIdentityManyBodyParallel} and
  \eqref{EQgIdentityManyBodyPerpendicular} remain valid
  \cite{sammueller2023supplementary}.}
 Our equilibrium theory requires proper thermal averaging for the
 presented identities to hold. A trivial counterexample is a
 precipitous temperature quench where the distribution of microsctates
 remains instantaneously intact, but $\beta$ has acquired a new
 value. Then the sum rule \eqref{EQgIdentity} is immediately violated,
 due to the respective scaling of the correlators $\nabla\nabla g(r)$,
 ${\sf g}_\gradf(r)$, and ${\sf g}_\ff(r)$ with powers $\beta^0$,
 $\beta^1$, and $\beta^2$.}

In conclusion, we have formulated and tested a systematic two-body
correlation framework based on invariance against an intrinsic
symmetry of thermal many-body systems. 
Formal similarities {\ms exist} with sum rules for interfacial
Hamiltonians \cite{mikheev1991}, as used for studies of wetting
\cite{squarcini2022}, {\ms and with Takahashi-Ward identities
  \cite{ward1950, takahashi1957} of quantum field theory.}  
{\ms Future work could relate to} the effective temperature
\cite{saw2022configurationalTemperature}, to one-dimensional systems
\cite{beenakker2022,flack2022,montero2019}, the structure of crystals
\cite{walz2010, haering2015, lin2021}, gels
\cite{saw2009,saw2011,lindquist2016}, {\ms glasses
  \cite{nandi2017,janssen2018,luo2022} and the hexatic phase
  \cite{bernard2011}}, to force-sampling simulation techniques
\cite{rotenberg2020, delasheras2018forceSampling, borgis2013}, and to
force-based classical \cite{tschopp2022forceDFT,
  sammueller2022forceDFT} and quantum density functional theory
\cite{ullrich2006, 
tchenkoue2019}.
{\ms Testing sum rules in charged systems is valuable, but can be
  technically subtle \cite{falcongonzales2020}.
Connections to three-point \cite{luo2022} and four-point
\cite{zhang2020,sing2023} correlation functions are interesting, as
for a simple fluid ${\sf g}_\ff(r)$ is given via two position
integrals over the four-body density.
We have checked that for molecular liquids the general force
correlation sum rules \eqref{EQsumRule2generic} and
\eqref{EQsumRuleFeFeDistinct} remain valid upon supplementing the
dependence on positions $\rv, \rv'$ with dependence on the molecular
orientational degrees of freedom; an analogous structure holds for
mixtures of different components.  Deriving torque correlation sum
rules requires using a local version of the Noether rotational
invariance \cite{hermann2021noether}.  }

{\ms A particularly exciting prospect is to apply the general identity
  \eqref{EQsumRuleFeFeDistinct} to the study of interfacial phenomena
  \cite{henderson1984, henderson1985, evans1990, upton1998,
    coe2022water, triezenberg1972}, where the connections with the
  existing body of sum rules \cite{hansen2013, henderson1984,
    henderson1985, henderson1992, evans1992, evans1990, upton1998,
    hirschfelder1960, triezenberg1972} and the constraints that follow
  on the allowed correlation function structure at complete drying
  \cite{evans2019pnas, coe2022prl, coe2023} and wetting transitions
  are worth exploring.}
Besides measurements of $g(r)$ \cite{royall2007gofr, thorneywork2014,
  statt2016, ramirez2006, yarnell1973, salmon2006, 
  carvalho2022}, position-resolved forces have recently become
accessible by direct imaging in colloidal systems \cite{dong2022},
which can facilitate experimental investigations of Noether
correlators.

\acknowledgments We thank Andrew O.\ Parry for pointing out
Ref.~\cite{mikheev1991} to us, and we are grateful to him, Gerhard
Jung, Atreyee Banerjee, Liesbeth Janssen, and Marjolein Dijkstra for
useful discussions.


\clearpage

\begin{widetext}
{\bf \large \noindent Supplementary Information\vspace{0.5mm}\\
 Noether-Constrained Correlations in Equilibrium Liquids
}\vspace{2mm}

\noindent Florian Samm\"uller, Sophie Hermann, Daniel de las Heras, 
and Matthias Schmidt\\
{\it Theoretische Physik II, Physikalisches Institut, 
  Universit{\"a}t Bayreuth, D-95447 Bayreuth, Germany}\\
\noindent (Dated: 
7 June 2023, 
\href{https://www.mschmidt.uni-bayreuth.de}
     {www.mschmidt.uni-bayreuth.de})\\
\vspace{5mm}

\setcounter{figure}{2} 

\begin{figure*}[h]
    \includegraphics[width=0.99\linewidth]
                    {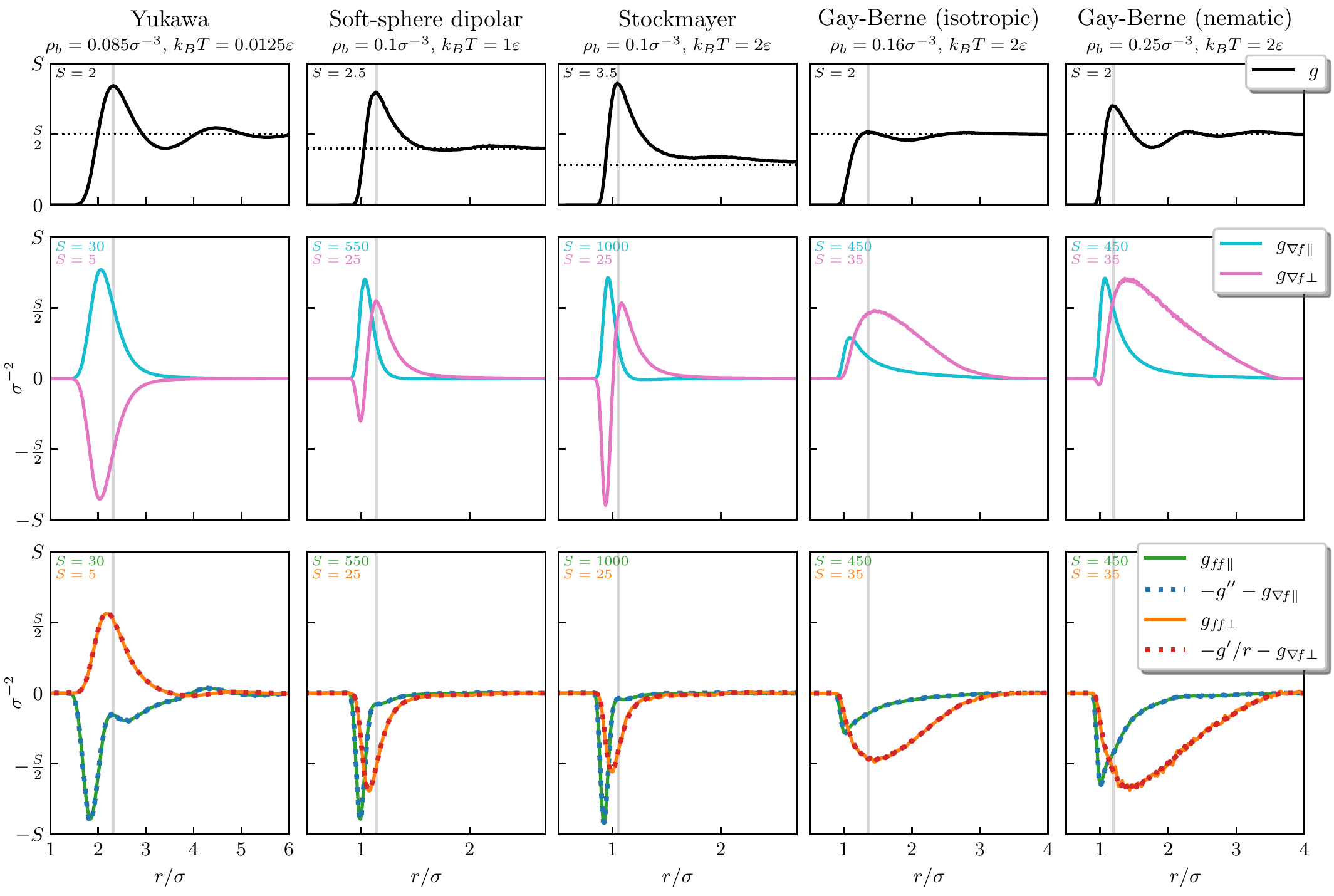}
\caption{Correlation functions analogous to Fig.~2 of the main text,
  but for the Yukawa liquid (first column), the soft-sphere dipolar
  fluid (second column), the Stockmayer fluid (third column), and the
  Gay-Berne model in the isotropic (fourth column) and nematic phase
  (fifth column). The results for the anisotropic models are obtained
  from canonical Monte Carlo simulations, and they are averaged over
  the microscopic orientations; the simulation box volume is
  $V=(20\sigma)^3$ and the long-ranged interactions are cut off at
  radial distance $10\sigma$. Shown are results for the pair
  correlation function $g(r)$ (first row) and for the radial
  ($\parallel$) and transversal ($\perp$) components of the two-body
  force-gradient correlator ${\sf g}_\gradf(r)$ (second row) and the
  force-force pair correlator ${\sf g}_\ff(r)$ (third row). The
  respective vertical scale factor $S$ is given in the top left corner
  of each panel and the scaled values for bulk density $\rho_b$ and
  temperature $T$ are indicated for each model fluid above the
  respective column.
  The results for the Yukawa liquid with inverse screening parameter
  $\kappa=2/\sigma$ are qualitatively similar to those of the WCA
  liquid (second column Fig.~2 of the main text) but here with much
  longer-ranged decay behaviour.  For identical dipolar strength
  $\mu/\sqrt{\epsilon\sigma^3}=2$ the results for $g_{\gradf\perp}(r)$
  for both the soft-sphere dipolar fluid \cite{stevens1995} and the
  Stockmayer fluid show strong signatures of chain formation, similar
  to the behaviour of the three-body gel former (fourth column of
  Fig.~2 of the main text).
  The Gay-Berne model (with parameters $\kappa = 3.8, \kappa' = 5$
  \cite{brown1998}) features positive-valued $g_{\gradf\perp}(r)$,
  which contrasts the behaviour of all other models and which we take
  to indicate interlocked arrangements of neighboring anisotropic
  molecules.
   \label{FIGsupplementary}
  }
\end{figure*}

\begin{figure*}[h]
    \includegraphics[height=11.5cm]{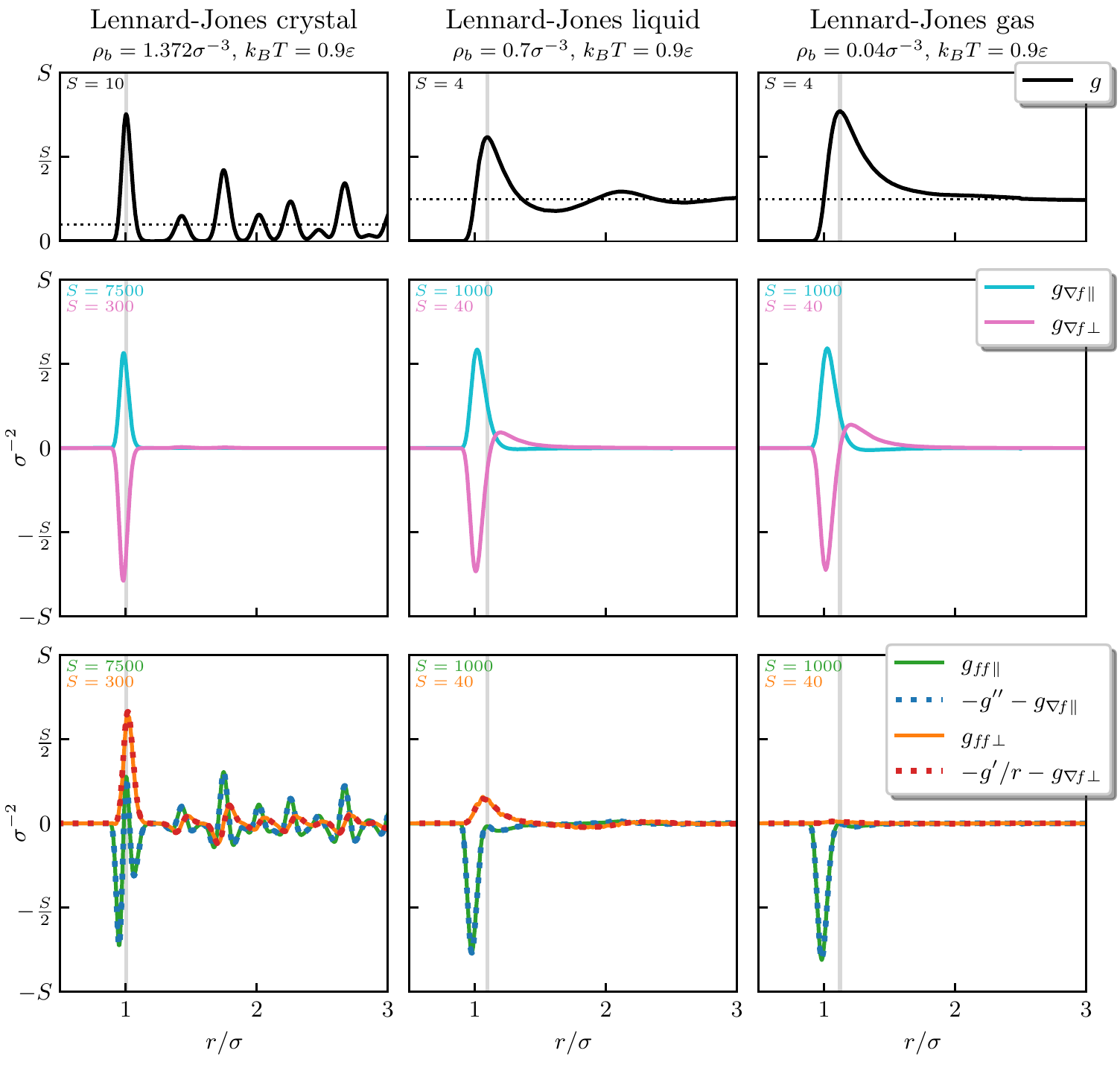}
\caption{
  Comparison of correlation functions for the LJ system in the fcc
  crystal phase (first colunm), the liquid (second column), and the
  gas phase (third column).
  Shown are the pair correlation function $g(r)$ (top row), the
  force-gradient correlator ${\sf g}_\gradf(r)$ (middle row), and the
  force-force correlator ${\sf g}_\ff(r)$ (bottom row). The plot style
  is analogous to Fig.~2 of the main text and to Fig.~3 of this SI.
  While the results for the gas and for the liquid carry the full
  structural two-body information, the correlators for the crystal are
  resolved only as a function of the radial distance $r$. This
  representation constitutes an average over global translations and
  rotations of the general inhomogeneous Noether sum rule, see
  Eq.~(11) of the main text with $|\rv-\rv'|$ kept fixed. The reduced
  Noether identities (13) and (14) continue to hold in this averaged
  sense, as is demonstrated by the data collapse in the lower left
  panel. This perfect agreement serves as an indirect indication of
  the validity of the more general Eq.~(11), which is applicable
  in the full inhomogeneous geometry.
   \label{FIGsupplSolid}
  }
\end{figure*}

\clearpage\end{widetext}


\begin{thebibliography}{31}

\bibitem{weisskopf1977}
  V. F. Weisskopf,
  {About Liquids,}
  \href{https://doi.org/10.1111/j.2164-0947.1977.tb02959.x}
       {Trans. N. Y. Acad. Sci. {\bf 38}, 202 (1977).}

\bibitem{evans2019physicsToday}
  R. Evans, D. Frenkel, and M. Dijkstra, 
  {From simple liquids to colloids and soft matter},
  \href{https://doi.org/10.1063/PT.3.4135}
     {Phys. Today {\bf 72}, 38 (2019).}

\bibitem{barker1976}
  J. A. Barker and D. Henderson,
  {What is ``liquid''? Understanding the states of matter,}
  \href{https://doi.org/10.1103/RevModPhys.48.587}
       {Rev. Mod. Phys. {\bf 48}, 587 (1976).}


\bibitem{hansen2013} 
  J.~P. Hansen and I.~R. McDonald, {\it Theory of
  Simple Liquids}, 4th ed.\  (Academic Press, London, 2013). 

\bibitem{evans1979} R. Evans,
  {The nature of the liquid-vapour interface and other topics 
    in the statistical mechanics of
  non-uniform, classical fluids,} 
  \href{https://doi.org/10.1080/00018737900101365}
       {Adv. Phys. {\bf 28}, 143 (1979).}

\bibitem{evans2016}
  R. Evans, M. Oettel, R. Roth, and G. Kahl, 
  {New developments in classical density functional theory,}
  \href{https://doi.org/10.1088/0953-8984/28/24/240401}
       {J. Phys.: Condens. Matter {\bf 28}, 240401 (2016).}

\bibitem{royall2007gofr}
  C. P. Royall, A. A. Louis, and H. Tanaka,
  {Measuring colloidal interactions with confocal microscopy,}
  \href{https://doi.org/10.1063/1.2755962}
       {J. Chem. Phys. {\bf 127}, 044507 (2007).}

\bibitem{thorneywork2014}
  A. L. Thorneywork, R. Roth, D. G. A. L. Aarts, and R. P. A. Dullens, 
  {Communication: Radial distribution functions in a two-dimensional binary 
    colloidal hard sphere system,}
  \href{http://dx.doi.org/10.1063/1.4872365}
       {J. Chem. Phys. {\bf 140}, 161106 (2014).}

\bibitem{statt2016}
  A. Statt, R. Pinchaipat, F. Turci, R. Evans, and C. P. Royall,
  {Direct observation in 3d of structural crossover in binary hard
    sphere mixtures,}
  \href{https://doi.org/10.1063/1.4945808}
       {J. Chem. Phys. {\bf 144}, 144506 (2016)}

{\ms \bibitem{ramirez2006}
  A. Ramirez-Saito, C. Bechinger, and J. L. Arauz-Lara,
  Optical microscopy measurement of pair correlation functions,
  \href{http://dx.doi.org/10.1103/PhysRevE.74.030401}
       {Phys. Rev. E {\bf 74}, 030401(R) (2006).}

\bibitem{yarnell1973}
  J. L. Yarnell, M. J. Katz, R. G. Wenzel, and S. H. Koenig,
  Structure factor and radial distribution function for liquid argon at 85K,
  \href{https://doi.org/10.1103/PhysRevA.7.2130}
       {Phys. Rev. A {\bf 7}, 2130 (1973).}

\bibitem{salmon2006}
  P. S. Salmon,
  Decay of the pair correlations and small-angle
  scattering for binary liquids and glasses,
  \href{http://dx.doi.org/10.1088/0953-8984/18/50/004}
       {J. Phys.: Condens. Matter {\bf 18}, 11443 (2006).}


\bibitem{carvalho2022}
  F. S. Carvalho, J. P. Braga,
  Partial radial distribution functions for a two-component glassy solid,
  GeSe, from scattering experimental data using an artificial 
  intelligence framework,
  \href{https://doi.org/10.1007/s00894-022-05055-5}
       {J. Molec. Modeling {\bf 28}, 99 (2022).}

\bibitem{dyre2016topicalReview}
  J. C. Dyre,
  Simple liquids' quasiuniversality and the hard--sphere paradigm,
  \href{https://doi.org/10.1088/0953-8984/28/32/323001}
       {J. Phys.: Condens. Matter {\bf 28}, 323001 (2016).}

}

\bibitem{evans1993decay}
  R. Evans, J. R. Henderson, D. C. Hoyle, A. O. Parry, and Z. A. Sabeur,
  {Asymptotic decay of liquid structure: oscillatory liquid-vapour 
  density profiles and the Fisher-Widom line,}
  \href{https://doi.org/10.1080/00268979300102621}
       {Mol. Phys. {\bf 80}, 755 (1993).}

\bibitem{evans1994decay}
  R. Evans, R. J. F. Leote de Carvalho, J. R. Henderson, and D. C. Hoyle,
  {Asymptotic decay of correlations in liquids and their mixtures,}
  \href{https://doi.org/10.1063/1.466920}
       {J. Chem. Phys. {\bf 100}, 591 (1994).}

{\ms \bibitem{dijkstra2000decay}
  M. Dijkstra and R. Evans,
  {A simulation study of the decay of the pair correlation function in 
  simple fluids,}
  \href{https://doi.org/10.1063/1.480598}
       {J. Chem. Phys. {\bf 112}, 1449 (2000).}}

\bibitem{grodon2004decay}
  C. Grodon, M. Dijkstra, R. Evans, and R. Roth,
  {Decay of correlation functions in hard-sphere mixtures: Structural crossover,}
  \href{https://doi.org/10.1063/1.1798057}
       {J. Chem. Phys. {\bf 121}, 7869 (2004).}

\bibitem{cats2021decay}
  P. Cats, R. Evans, A. H\"artel, and R. van Roij,
  {Primitive model electrolytes in the near and far field: Decay lengths 
  from DFT and simulations,}
  \href{https://doi.org/10.1063/5.0039619}
       {J. Chem. Phys. {\bf 154}, 124504 (2021).}

\bibitem{noether1918}
  E. Noether,
  {Invariante Variationsprobleme,}
  \href{https://gdz.sub.uni-goettingen.de/download/pdf/PPN252457811_1918/LOG_0022.pdf}
       {Nachr. d. K\"onig. Gesellsch. d. Wiss. zu G\"ottingen, 
  Math.-Phys. Klasse, {\bf 235}, 183 (1918).}
%
  English translation by M. A. Tavel: Invariant variation
  problems. Transp. Theo. Stat.  Phys. {\bf 1}, 186 (1971); for a
  version in modern typesetting see: Frank Y. Wang,
  \href{http://arxiv.org/abs/physics/0503066v3}{arXiv:physics/0503066v3}
  (2018).

\bibitem{byers1998} 
  N. Byers,
  {E.\ Noether's discovery of the deep connection between
  symmetries and conservation laws,}
  \href{https://doi.org/10.48550/arXiv.physics/9807044}
       {arXiv:physics/9807044 (1998)}.

\bibitem{lezcano2018stochastic}
  A. G. Lezcano and A. C. M. de Oca, 
  {A stochastic version of the Noether theorem,}
  \href{https://doi.org/10.1007/s10701-018-0174-z}
       {Found. Phys. \textbf{48}, 726 (2018).}

\bibitem{baez2013markov}
  J. C. Baez and B. Fong, 
  {A Noether theorem for Markov processes,}
  \href{http://dx.doi.org/10.1063/1.4773921}
       {J. Math. Phys. \textbf{54}, 013301 (2013).}

\bibitem{marvian2014quantum}
  I. Marvian and R. W. Spekkens, 
  {Extending Noether’s theorem by quantifying the asymmetry of quantum states,}
  \href{https://doi.org/10.1038/ncomms4821}
       {Nat. Commun. \textbf{5}, 3821 (2014).}

\bibitem{sasa2016}
   S. Sasa and Y. Yokokura,
   {Thermodynamic entropy as a Noether invariant,}
   \href{http://dx.doi.org/10.1103/PhysRevLett.116.140601}
   {Phys. Rev. Lett. {\bf 116}, 140601 (2016).}
   
\bibitem{sasa2019} 
  S. Sasa, S. Sugiura, and Y. Yokokura, 
  {Thermodynamical path integral and emergent symmetry,}
  \href{https://doi.org/10.1103/PhysRevE.99.022109}
       {Phys. Rev. E \textbf{99}, 022109 (2019).}

\bibitem{sasa2020}
  Y. Minami and S. Sasa, 
  {Thermodynamic entropy as a Noether invariant in a Langevin equation,}
  \href{https://doi.org/10.1088/1742-5468/ab5b8b}
       {J. Stat. Mech. \textbf{2020}, 013213 (2020).}

\bibitem{revzen1970}
  M. Revzen,
  {Functional integrals in statistical physics,}
  \href{https://doi.org/10.1119/1.1976414}
       {Am. J. Phys. {\bf 38}, 611 (1970).}


{\ms \bibitem{budkov2022}
  Y. A. Budkov and A. L. Kolesnikov,
  Modified Poisson-Boltzmann equations and macroscopic
  forces in inhomogeneous ionic fluids,
  \href{https://doi.org/10.1088/1742-5468/ac6a5b}
       {J. Stat. Mech. {\bf 2022}, 053205 (2022).}

\bibitem{brandyshev2023}
  P. E. Brandyshev, Y. A. Budkov,
  Noether's second theorem and covariant field theory of mechanical
  stresses in inhomogeneous ionic fluids,
  \href{https://doi.org/10.1063/5.0148466}
       {J. Chem. Phys. {\bf 158}, 174114 (2023).}
}

\bibitem{hermann2021noether}
  S. Hermann and M. Schmidt, 
  {Noether's theorem in statistical mechanics,}
  \href{https://doi.org/10.1038/s42005-021-00669-2}
       {Commun. Phys.  {\bf  4}, 176 (2021).}

\bibitem{hermann2022topicalReview}
  S. Hermann and M. Schmidt, 
  {Why Noether's theorem applies to statistical mechanics,}
   \href{https://doi.org/10.1088/1361-648X/ac5b47}
        {J. Phys.: Condens. Matter 
          {\bf 34}, 213001 (2022) (Topical Review).}

\bibitem{hermann2022variance}
  S. Hermann and M. Schmidt, 
  {Variance of fluctuations from Noether invariance},
  \href{https://doi.org/10.1038/s42005-022-01046-3}
       {Commun. Phys. {\bf 5}, 276 (2022).}
       
\bibitem{hermann2022quantum}
  S. Hermann and M. Schmidt, 
  {Force balance in thermal quantum many-body systems from Noether's theorem,}
   \href{https://doi.org/10.1088/1751-8121/aca12d}
        {J. Phys. A: Math. Theor.  {\bf 55}, 464003 (2022).}

\bibitem{tschopp2022forceDFT}
  S. M. Tschopp, F. Samm\"uller, S. Hermann, M. Schmidt, and J.~M. Brader,
  {Force density functional theory in- and out-of-equilibrium},
  \href{https://doi.org/10.1103/PhysRevE.106.014115}
  {Phys. Rev. E {\bf 106}, 014115 (2022).}
  

{\ms

\bibitem{henderson1992}
  J. R. Henderson,
  ``Statistical mechanical sum rules,'' 
  in Fundamentals of Inhomogeneous Fluids, 
  edited by D. Henderson (Dekker, New York, 1992).

\bibitem{evans1992}
  R. Evans,
  ``Density functionals in the theory of non-uniform fluids,'' 
  in Fundamentals of Inhomogeneous Fluids, 
  edited by D. Henderson (Dekker, New York, 1992).

\bibitem{upton1998}
  P. J. Upton,
  {Fluids against hard walls and surface critical behavior,}
  \href{https://doi.org/10.1103/PhysRevLett.81.2300}
       {Phys. Rev. Lett. {\bf 81}, 2300 (1998).}

\bibitem{evans1990}
  R. Evans and A. O. Parry,
  Liquids at interfaces: what can a theorist contribute?
  \href{https://doi.org/10.1088/0953-8984/2/S/003}
       {J. Phys.: Condens. Matter {\bf 2}, SA15 (1990).}

\bibitem{henderson1984}
  J. R. Henderson and F. van Swol,
  On the interface between a fluid and a planar wall: theory and
  simulations of a hard sphere fluid at a hard wall,
  \href{https://doi.org/10.1080/00268978400100651}
       {Mol. Phys. {\bf 51}, 991 (1984).}

\bibitem{henderson1985}
  J. R. Henderson and F. van Swol,
  On the approach to complete wetting by gas at a liquid-wall
  interface,
  \href{https://doi.org/10.1080/00268978500103081}
       {Mol. Phys. {\bf 56}, 1313 (1985).}

\bibitem{hirschfelder1960}
  J. O. Hirschfelder,
  Classical and quantum mechanical hypervirial theorems,
  \href{https://doi.org/10.1063/1.1731427}
       {J. Chem. Phys. {\bf 33}, 1462 (1960).}
  
}

\bibitem{triezenberg1972}
  D. G. Triezenberg and R. Zwanzig, 
  {Fluctuation theory of surface tension,}
  Phys. Rev. Lett. \textbf{28}, 1183 (1972).




\bibitem{goldstein2002} 
  H. Goldstein,  C. Poole, and J. Safko,
  {\it Classical Mechanics}
  (Addison-Wesley, New York, 2002).
  Our generator $\cal G$ is given as $F_2$ in their notation.
  
\bibitem{schmidt2022rmp}
  M. Schmidt, 
  {Power functional theory for many-body dynamics},
  \href{https://doi.org/10.1103/RevModPhys.94.015007}
       {Rev. Mod. Phys. {\bf 94}, 015007 (2022).}

{\ms
\bibitem{zhao2011}
  S. Zhao, R. Ramirez, R. Vuilleumier, D. Borgis,
  Molecular density functional theory of solvation: From polar solvents to water,
  \href{https://doi.org/10.1063/1.3589142}
       {J. Chem. Phys. {\bf 134}, 194102 (2011).}

\bibitem{jeanmairet2013jpcl}
  G. Jeanmairet, M. Levesque, R. Vuilleumier, and D. Borgis,
  Molecular density functional theory of water,
  \href{https://doi.org/10.1021/jz301956b}
       {J. Phys. Chem. Lett. {\bf 4} 619 (2013).}


\bibitem{luukkonen2020}
  S. Luukkonen, M. Levesque, L. Belloni, and D. Borgis,
  Hydration free energies and solvation structures with molecular density 
  functional theory in the hypernetted chain approximation,
  \href{https://doi.org/10.1063/1.5142651}
       {J. Chem. Phys. {\bf 152}, 064110 (2020).}
}

\bibitem{sammueller2021}
  F. Samm\"uller and M. Schmidt, 
  {Adaptive Brownian dynamics},
  \href{https://doi.org/10.1063/5.0062396}
       {J. Chem. Phys. {\bf 155}, 134107 (2021).}

\bibitem{molinero2009}
  V. Molinero and E. B. Moore,
  {Water modeled as an intermediate element between carbon and silicon,}
  \href{https://doi.org/10.1021/jp805227c}
       {J. Phys. Chem. B {\bf 113}, 4008 (2009).}

\bibitem{coe2022water}
  M. K. Coe, R. Evans, and N. B. Wilding,
  {The coexistence curve and surface tension of a monatomic water model,}
  \href{https://doi.org/10.1063/5.0085252}
       {J. Chem. Phys. {\bf 156}, 154505 (2022).}

\bibitem{saw2009}
  S. Saw, N. L. Ellegaard, W. Kob, and S. Sastry, 
  {Structural relaxation of a gel modeled by three body interactions,}
  \href{https://doi.org/10.1103/PhysRevLett.103.248305}
       {Phys. Rev. Lett. {\bf 103}, 248305 (2009).}
  
\bibitem{saw2011}
  S. Saw, N. L. Ellegaard, W. Kob, and S. Sastry,
  {Computer simulation study of the phase behavior and structural
    relaxation in a gel-former modeled by three-body interactions,}
  \href{https://doi.org/10.1063/1.3578176}
       {J. Chem. Phys. {\bf 134}, 164506 (2011).}

\bibitem{sammueller2022gel}
  F. Samm\"uller, D. de las Heras, and M. Schmidt,
  {Inhomogeneous steady shear dynamics of a three-body colloidal gel former,}
  \href{https://doi.org/10.1063/5.0130655}
       {\ms J. Chem. Phys.  {\bf 158}, 054908 (2023)}.

{\ms
\bibitem{sammueller2023supplementary} See Supplementary Information
  for simulation results of the Noether correlation functions for the
  Yukawa, soft-sphere dipolar, Stockmayer, and Gay-Berne models
  (Fig.~3), as well as results for the gas, liquid, and crystal phase
  of the Lennard-Jones model (Fig.~4) .

\bibitem{teixeira2000topRev}
  P. I. C. Teixeira, J. M. Tavares, and M. M. Telo da Gama,
  Review Article: The effect of dipolar forces on the structure and thermodynamics
  of classical fluids,
  \href{https://doi.org/10.1088/0953-8984/12/33/201}
       {J. Phys.: Condens. Matter {\bf 12}, R411 (2000).}

\bibitem{klapp2005topRev}
  S. H. L. Klapp,
  Topical Review: Dipolar fluids under external perturbations,
  \href{https://doi.org/10.1088/0953-8984/17/15/R02}
       {J. Phys.: Condens. Matter {\bf 17}, R525 (2005).}

\bibitem{stevens1995}
  M. J. Stevens and G. S. Grest,
  Structure of soft-sphere dipolar fluids,
  \href{https://doi.org/10.1103/PhysRevE.51.5962}
       {Phys. Rev. E {\bf 51}, 5962 (1995).}

\bibitem{tavares1999}
  J. M. Tavares, J. J. Weis, and M. M. Telo da Gama,
  Strongly dipolar fluids at low densities compared to living polymers,
  \href{https://doi.org/10.1103/PhysRevE.59.4388}
       {Phys. Rev. E {\bf 59}, 4388 (1999).}

\bibitem{allen2019topRev}
  M. P. Allen,
  Topical Review: Molecular simulation of liquid crystals
  \href{https://doi.org/10.1080/00268976.2019.1612957}
       {Mol. Phys. {\bf 117}, 2391 (2019).}

\bibitem{gay1981}
  J. G. Gay and B. J. Berne,
  Modification of the overlap potential to mimic a linear site-site potential
  \href{https://doi.org/10.1063/1.441483}
       {J. Chem. Phys. {\bf 74}, 3316 (1981).}

\bibitem{brown1998}
  J. T. Brown, M. P. Allen, E. Martín del Río, and E. de Miguel,
  Effects of elongation on the phase behavior of the Gay-Berne fluid,
  \href{https://doi.org/10.1103/PhysRevE.57.6685}
       {Phys. Rev. E {\bf 57}, 6685 (1998).}
}


\bibitem{mikheev1991}
  L. V. Mikhheev and J. D. Weeks,
  {Sum rules for interface Hamiltonians,}
  \href{https://doi.org/10.1016/0378-4371(91)90192-F}
       {Physica A {\bf 177}, 495 (1991).}

\bibitem{squarcini2022}
  A. Squarcini, J. M. Romero-Enrique, and A. O. Parry,
  {Casimir Contribution to the Interfacial Hamiltonian for 3D Wetting,}
  \href{https://doi.org/10.1103/PhysRevLett.128.195701}
       {Phys. Rev. Lett. {\bf 128}, 195701 (2022).}

{\ms 
\bibitem{ward1950}
  J. C. Ward,
  An identity in quantum electrodynamics,
  \href{https://doi.org/10.1103/PhysRev.78.182}
       {Phys. Rev. {\bf 78}, 182 (1950).}

\bibitem{takahashi1957}
  Y. Takahashi,
  On the generalized Ward identity,
  \href{https://doi.org/10.1007/BF02832514}
       {Il Nuovo Cimento {\bf 6}, 371 (1957).}
}


\bibitem{saw2022configurationalTemperature}
  S. Saw, L. Costigliola, and J. C. Dyre,
  {Configurational temperature in active-matter models. 
    I. Lines of invariant physics in the phase diagram of the Ornstein-Uhlenbeck model},
  \href{https://doi.org/10.1103/PhysRevE.107.024609}
       {\ms Phys. Rev. E {\bf 107}, 024609 (2023).}

\bibitem{beenakker2022}
  C. W. J. Beenakker,
  {Pair correlation function of the one-dimensional Riesz gas},
  \href{https://doi.org/10.1103/PhysRevResearch.5.013152}
       {\ms Phys. Rev. Research {\bf 5}, 013152 (2023).}

\bibitem{flack2022}
  A. Flack, S. N. Majumdar, and G. Schehr,
  {An exact formula for the variance of linear statistics in the
    one-dimensional jellium model},
  \href{https://doi.org/10.1088/1751-8121/acb86a}
       {\ms J. Phys. A: Math. Theor. {\bf 56}, 105002 (2023).}

{\ms
\bibitem{montero2019}
  A. M. Montero and A. Santos,
  Triangle-well and ramp interactions in one-dimensional fluids: 
  a fully analytic exact solution,
  \href{https://doi.org/10.1007/s10955-019-02255-x}
       {J. Stat. Phys. {\bf 175}, 269 (2019).}
}

\bibitem{walz2010}
  C. Walz and M. Fuchs, 
  {Displacement field and elastic constants in nonideal crystals,}
  \href{https://doi.org/10.1103/physrevb.81.134110}
       {Phys. Rev. B {\bf 81}, 134110 (2010).}

\bibitem{haering2015}
  J. M. H\"aring, C. Walz, G. Szamel, and M. Fuchs,
  {Coarse-grained density and compressibility of nonideal crystals: 
    General theory and an application to cluster crystals,}
  \href{https://doi.org/10.1103/physrevb.92.184103}
       {Phys. Rev. B {\bf 92}, 184103 (2015).}

\bibitem{lin2021}
  S.-C. Lin, M. Oettel, J. M. H\"aring, R. Haussmann, M.~Fuchs, and G. Kahl,
  {Direct correlation function of a crystalline solid,}
  \href{https://doi.org/10.1103/PhysRevLett.127.085501}
       {Phys. Rev. Lett. {\bf 127}, 085501 (2021).}

\bibitem{lindquist2016}
  B. A. Lindquist, R. B. Jadrich, D. J. Milliron, and T.~M. Truskett,
  {On the formation of equilibrium gels via a macroscopic bond limitation,}
  \href{https://doi.org/10.1063/1.4960773}
       {J. Chem. Phys. {\bf 145}, 074906 (2016).}

{\ms
\bibitem{nandi2017}
  M. K. Nandi, A. Banerjee, C. Dasgupta, S. M. Bhattacharyya,
  Role of the pair correlation function in the dynamical transition 
  predicted by mode coupling theory,
  \href{https://doi.org/10.1103/PhysRevLett.119.265502}
       {Phys. Rev. Lett. {\bf 119}, 265502 (2017).}

\bibitem{janssen2018}
  L. M. C. Janssen,
  Mode-coupling theory of the glass transition: A primer,
  \href{https://doi.org/10.3389/fphy.2018.00097}
       {Front. Phys. {\bf 6}, 97 (2018).}

\bibitem{luo2022}
  C. Luo, J. F. Robinson, I. Pihlajamaa, V. E. Debets, C. P. Royall, 
  and L. M. C. Janssen,
  Many-body correlations are non-negligible in both fragile and
  strong glassformers,
  \href{https://doi.org/10.1103/PhysRevLett.129.145501}
       {Phys. Rev. Lett. {\bf 129}, 145501 (2022).}


\bibitem{bernard2011}
  E. P. Bernard and W. Krauth.
  Two-Step melting in two dimensions: first-order liquid-hexatic transition,
  \href{https://doi.org/10.1103/PhysRevLett.107.155704}
       {Phys. Rev. Lett. {\bf 107}, 155704 (2011).}


}


\bibitem{rotenberg2020}
  B. Rotenberg,
  {Use the force! Reduced variance estimators for densities,
    radial distribution functions, and local mobilities in
    molecular simulations,}
  \href{https://doi.org/10.1063/5.0029113}
       {J. Chem. Phys. {\bf 153}, 150902 (2020).}

\bibitem{delasheras2018forceSampling}
  D. de las Heras and M. Schmidt,
  {Better than counting: Density profiles from force sampling,}
  \href{https://doi.org/10.1103/PhysRevLett.120.218001}
       {Phys. Rev. Lett. {\bf 120}, 218001 (2018).}

\bibitem{borgis2013}
  D. Borgis, R. Assaraf, B. Rotenberg, and R. Vuilleumier,
  {Computation of pair distribution functions and three-dimensional
    densities with a reduced variance principle,}
  \href{https://doi.org/10.1080/00268976.2013.838316}
       {Mol. Phys. {\bf 111}, 3486 (2013).}

\bibitem{sammueller2022forceDFT}
  F. Samm\"uller, S. Hermann, and M. Schmidt,
 {Comparative study of force-based classical density functional theory},
  \href{https://doi.org/10.1103/PhysRevE.107.034109}
       {\ms  Phys. Rev. E. {\bf 107}, 034109 (2023).}



\bibitem{ullrich2006}
  C. A. Ullrich and I. V. Tokatly,
  {Nonadiabatic electron dynamics in time-dependent density-functional theory},
  \href{https://doi.org/10.1103/PhysRevB.73.235102}
       {Phys. Rev. B {\bf 73}, 235102 (2006).}

\bibitem{tchenkoue2019}
  M.~M. Tchenkoue, M. Penz, I. Theophilou, M. Ruggenthaler, and A. Rubio,
  {Force balance approach for advanced approximations in 
  density functional theories,}
  \href{https://doi.org/10.1063/1.5123608}
       {J. Chem. Phys. {\bf 151}, 154107 (2019).}




{\ms


\bibitem{falcongonzales2020}
  J. M. Falc\'on-Gonz\'alez, C. Contreras-Aburto, M. Lara-Pe\~na,
  M. Heinen, C. Avenda\~no, A. Gil-Villegas, and R. Casta\~neda-Priego,
  Assessment of the Wolf method using the Stillinger-Lovett sum rules:
  from strong electrolytes to weakly charged colloidal dispersions,
  \href{https://doi.org/10.1063/5.0033561}
       {J. Chem. Phys. {\bf 153}, 234901 (2020).}

\bibitem{zhang2020}
  Z. Zhang and W. Kob,
  Revealing the three-dimensional structure of liquids using
  four-point correlation functions,
  \href{https://doi.org/10.1073/pnas.2005638117}
       {PNAS {\bf 117}, 14032 (2020).}
       
\bibitem{sing2023}
  N. Singh, Z. Zhang, A. K. Sood, W. Kob, and R. Ganapathy,
  Intermediate-range order governs dynamics in dense colloidal liquids,
  \href{https://doi.org/10.1073/pnas.2300923120}
       {Proc. Nat. Acad. Sci. {\bf 120}, e2300923120 (2023).}


\bibitem{evans2019pnas}
  R. Evans, M. C. Stewart, and N. B. Wilding,  
  {A unified description of hydrophilic and superhydrophobic surfaces in 
  terms of the wetting and drying transitions of liquids},
  \href{https://doi.org/10.1073/pnas.1913587116}
       {Proc. Nat. Acad. Sci. {\bf 116}, 23901 (2019).}


\bibitem{coe2022prl}
  M. K.  Coe, R. Evans, and N. B. Wilding,
  {Density depletion and enhanced fluctuations in water near 
    hydrophobic solutes: identifying the underlying physics,}
  \href{https://doi.org/10.1103/PhysRevLett.128.045501}
       {Phys. Rev. Lett. {\bf 128}, 045501 (2022).}

\bibitem{coe2023}
  M. K.  Coe, R. Evans, and N. B. Wilding,
  {Understanding the physics of hydrophobic solvation,}
  \href{https://doi.org/10.1063/5.0134060}
       {J. Chem. Phys. {\bf 158}, 034508 (2023).}

}

\bibitem{dong2022}
  J. Dong, F. Turci, R. L. Jack, M. A. Faers, and C. P. Royall,
  {Direct imaging of contacts and forces in colloidal gels,}
  \href{https://doi.org/10.1063/5.0089276}
       {J. Chem. Phys. {\bf 156}, 214907 (2022).}



\end{thebibliography}
\end{document}